\documentclass[11pt,a4]{article}
\usepackage{times,fancyhdr}
\usepackage[dvips]{graphicx}
\usepackage{array}
\usepackage{amssymb}
\usepackage[small,bf]{caption}
\usepackage{color}
\usepackage{fancyhdr}
\usepackage{natbib}
\usepackage{setspace}
\usepackage{framed}
\usepackage{xcolor}
\usepackage{verbatim,framed}
\usepackage[yyyymmdd,hhmmss]{datetime}
\fancyheadoffset[R]{0pt}
\fancyfootoffset[R]{0pt}
\renewcommand{\headrulewidth}{0.4pt}
\renewcommand{\footrulewidth}{0.4pt}
\definecolor{orange}{rgb}{1.0,0.5,0.0}
\definecolor{aqgr}  {rgb}{0.0,1.0,0.6} 
\definecolor{viol}  {rgb}{0.8,0.6,0.8}
\definecolor{figdr} {rgb}{1.0,1.0,1.0} 
\definecolor{coldr} {rgb}{1.0,0.0,1.0} 
\definecolor{colwp} {rgb}{1.0,0.8,0.0} 
\definecolor{colok} {rgb}{0.8,1.0,0.5} 
\definecolor{colmb} {rgb}{1.0,1.0,1.0}

\newcolumntype{C}[1]{>{\centering\let\newline\\\arraybackslash\hspace{0pt}}m{#1}}
\def\nr{\rm \color{black}}

\title{\bfseries{\textsc{Epigenetic Tracking: \\
a Model for \\
Multicellular Biology}}}

\author{Alessandro Fontana \\
\mbox{}\\
email address: fontalex00@gmail.com}

\begin{document}
\maketitle
   
\clubpenalty=20000
\widowpenalty=20000

\begin{abstract}
Epigenetic Tracking is a mathematical model of biological cells, originally conceived to study embryonic development. Computer simulations proved the capacity of the model to generate complex 3-dimensional cellular structures, and the potential to reproduce the complexity typical of living beings. The most distinctive feature of this model is the presence in the body of a homogeneous distribution of stem cells, which are dinamically and continuously created during development from non-stem cells and reside in niches. Embryonic stem cells orchestrate early developmental events, adult stem cells direct late developmental and regeneration events, ageing stem cells cause ageing and cancer stem cells are responsible for cancer growth. The conceptual backbone provided by Epigenetic Tracking brings together a wide range of biological phenomena: for this reason, we think it can be proposed as a general model for multicellular biology. Despite, or perhaps due to its theoretical origin, the model allowed us to make predictions relevant to very diverse fields of biology, such as transposable elements, and cancer-related patterns of gene mutations. This paper contains a summary of the model and its implications. 
\end{abstract}

\section{Introduction}

This work is concerned with a biological model called \textit{Epigenetic Tracking (ET)}, described in \cite{Fontana08, Fontana09, Fontana10a, Fontana10b, Fontana10c, Fontana12b, Fontana13c, Fontana13d, Fontana14a}. While the study of life sciences often relies on a bottom-up method, that tries to infer general rules starting from genes and proteins, our work is informed by a top-down approach, that proceeds from high-level abstractions towards a proposal for biological, low-level molecular processes. As a consequence our model may contain ingredients not necessarily adherent to current knowledge, but which can become a suggestion for biologists to look into new, previously unexplored directions.

The capacity of the model to generate complex cellular structures was proved through computer simulations. If we interpret such structures as a metaphor for biological organisms, we can see in this endeavour the potential to reproduce the complexity typical of living beings. Furthermore, ET is able to interpret with a unified framework a wide array of biological phenomena, such as development, the presence of junk DNA, ageing and cancer: for this reason, we think it can be proposed as a general model for multicellular biology. This paper is divided into six sections: this (first) section is the introduction; section 2 describes ET as a model of development; section 3 is dedicated to the evo-devo process; section 4 deals with ageing; section 5 explores the topic of cancer; section 6 draws the conclusions and outlines future directions.

\section{Development}
\label{sec:cellular-model}

\begin{figure}[t] \begin{center} \hspace*{-0.75cm}
{\fboxrule=0.0mm\fboxsep=0mm\fbox{\includegraphics[width=18.00cm]{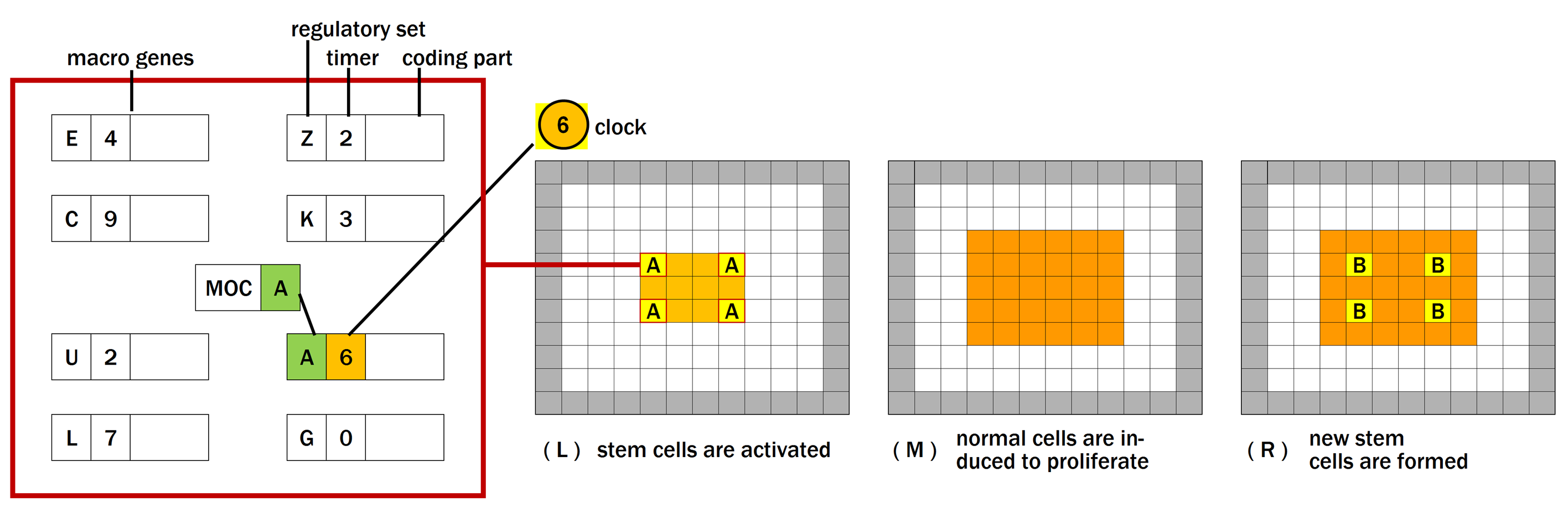}}}
\caption{Developmental events. When in a stem cell the mobile code matches with the regulatory set of a macro gene and the clock matches with the timer, the stem cell gets activated and performs actions encoded in the coding part of the gene. Upon activation, stem cells (with mobile code A, marked in yellow in panel L), can induce normal cells (in orange) to proliferate (panel M) or to differentiate (not shown). Other stem cells, with a different mobile code (B), are generated from normal cells at a subsequent stage (panel R). The mobile code is present also in normal cells, but it is active in stem cells only.}
\label{events}
\end{center} \end{figure}

\colorbox{colmb}{\textit{Hox genes}}are a family of regulatory genes, characterised by the presence of a DNA sequence called homeobox. In early embryogenesis, Hox gene regulation is achieved through the maternal factors present in the fertilised egg, gap and pair-rule genes \cite{Pearson05}. In sebsequent phases of development this role is taken over by \textit{morphogens}, a class of compounds (e.g., BMP, Hedgehog, Notch, Wnt) which spread from localised sources and form concentration gradients, used by cells to induce the expression of specific genes, through a characteristic ``morphogenetic code'' \cite{Hogan99}. 

\colorbox{colmb}{\textit{Growth factors}}are a class of molecules which during embryonic development act locally, either as paracrine or autocrine regulatory chemical messengers, as important regulators of cellular proliferation and differentiation \cite{Ruddon09}. It is hypothesised that each organ or organ part requires the action of a specific subset of growth factors. Each morphogen and growth factor is processed by a dedicated signalling pathway, that starts when a ligand binds a receptor on the cell surface, and ends in the nucleus, where the induced signal participates in the regulation of target genes. 

\colorbox{colmb}{In biological} development a pivotal role is known to be played by \textit{stem cells}, a class of cells found in most multicellular organisms. Embryonic stem cells are totipotent cells found in the inner cell mass of the blastocyst. Adult stem cells are pluripotent undifferentiated cells found in ``niches'' throughout the body, where a particular microenvironment is required to maintain the stem state \cite{Song04}. The question of when and how adult stem cells are formed remains an open issue in biology.

In ET development starts from a single cell and unfolds in a predefined number of \textit{developmental stages}, counted by a \textit{global clock} shared by all cells. Artificial bodies are composed of two categories of cells: \textit{normal cells} and \textit{stem cells}. Cells have a variable called \textit{mobile code (MOC)}, which can take different values in different cells (different mobile codes are shown with different letters). The mobile code is present also in normal cells, but it is active in stem cells only. 

\colorbox{colmb}{The mobile}code can be interpreted in biology as the set of master transcription factors present in a cell, which are responsible for the orchestration of developmental events. In our model the mobile code is hypothesised to be composed of Hox proteins (in nature this may be true in early embryogenesis, while at later stages the master role may be played also by other elements). The clock represents a further transcription factor, carrying a temporal condition for gene activation.

Stem cells direct development. In ET, this is achieved through the interaction of the mobile code and the clock with a set of macro genes (Fig.~\ref{events}). Each macro gene has a variable called \textit{regulatory set}, that can match with the mobile code, a variable called \textit{timer}, that can match with the clock, and a coding part. When a match occurs in a given stem cell, the cell gets activated and a specific \textit{developmental event} (encoded in the coding part of the gene) is orchestrated. Macro genes have also a \textit{switch} (not shown in the figure): when the switch is in the OFF position, the gene is inactive. Macro genes are hypothesised to correspond in biology to sets of genes co-regulated in space and time, involved in the production of growth factors (the key regulators of biological developmental events).

When stem cells are activated (Fig.~\ref{events}), they induce normal cells present in the volume nearby to proliferate or to differentiate. If no normal cells are present in the volume, they can be generated by stem cells through several rounds of asymmetric cell divisions (i.e. the stem cell produces another identical stem cell and a normal cell). Subsequently, some normal cells revert to stem cells, and receive new mobile codes. Some of these newly formed stem cells can become the centre of other developmental events at a later stage, and development can move ahead. Hence, in ET the normal-stem transition is a two-way street.

\begin{figure}[t] \begin{center} \hspace*{-0.0cm}
{\fboxrule=0.0mm\fboxsep=0mm\fbox{\includegraphics[width=16.00cm]{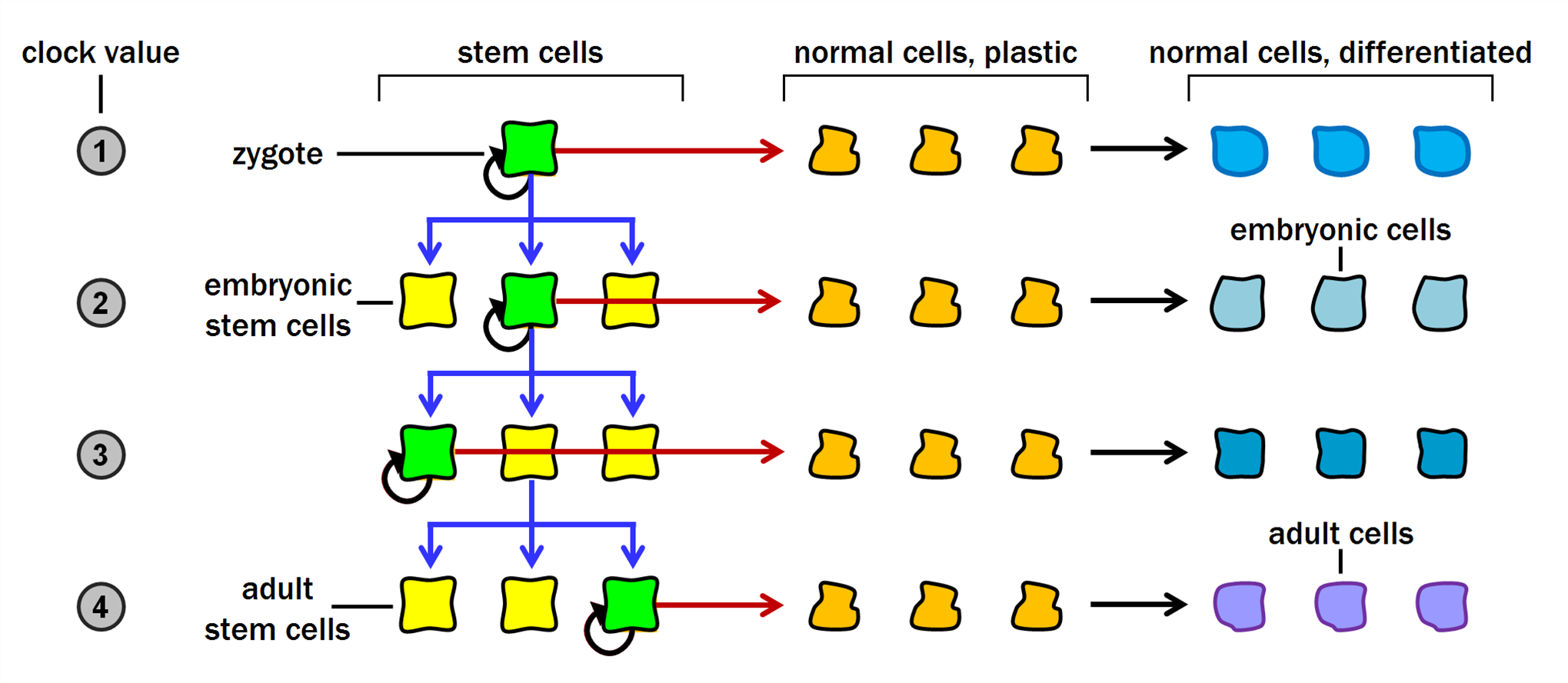}}}
\caption{Set of stem cells created during development. Stem cells marked in green become active during development. As a result, they create normal cells and induce their differentiation. In parallel, other stem cells are created.}
\label{biocells}
\end{center} \end{figure}

The set of stem cells and mobile codes generated during development has a hierarchical, or tree, structure (Fig.~\ref{biocells}). This feature appears consistent with the information we have on the set of biological stem cells involved in the generation of particular organs or systems, such as the hematopoietic system \cite{Reya01}. Our model provides also a means to bridge the conceptual gap between embryonic and adult stem cells: embryonic stem cells correspond to elements in the tree near the root, while adult stem cells correspond to the leaves of the tree. It also explains how adult stem cells are generated, with the mechanism of conversion from normal cells.

\begin{figure}[p] \begin{center} \hspace*{-0.0cm}
{\fboxrule=0.0mm\fboxsep=0mm\fbox{\includegraphics[width=15.80cm]{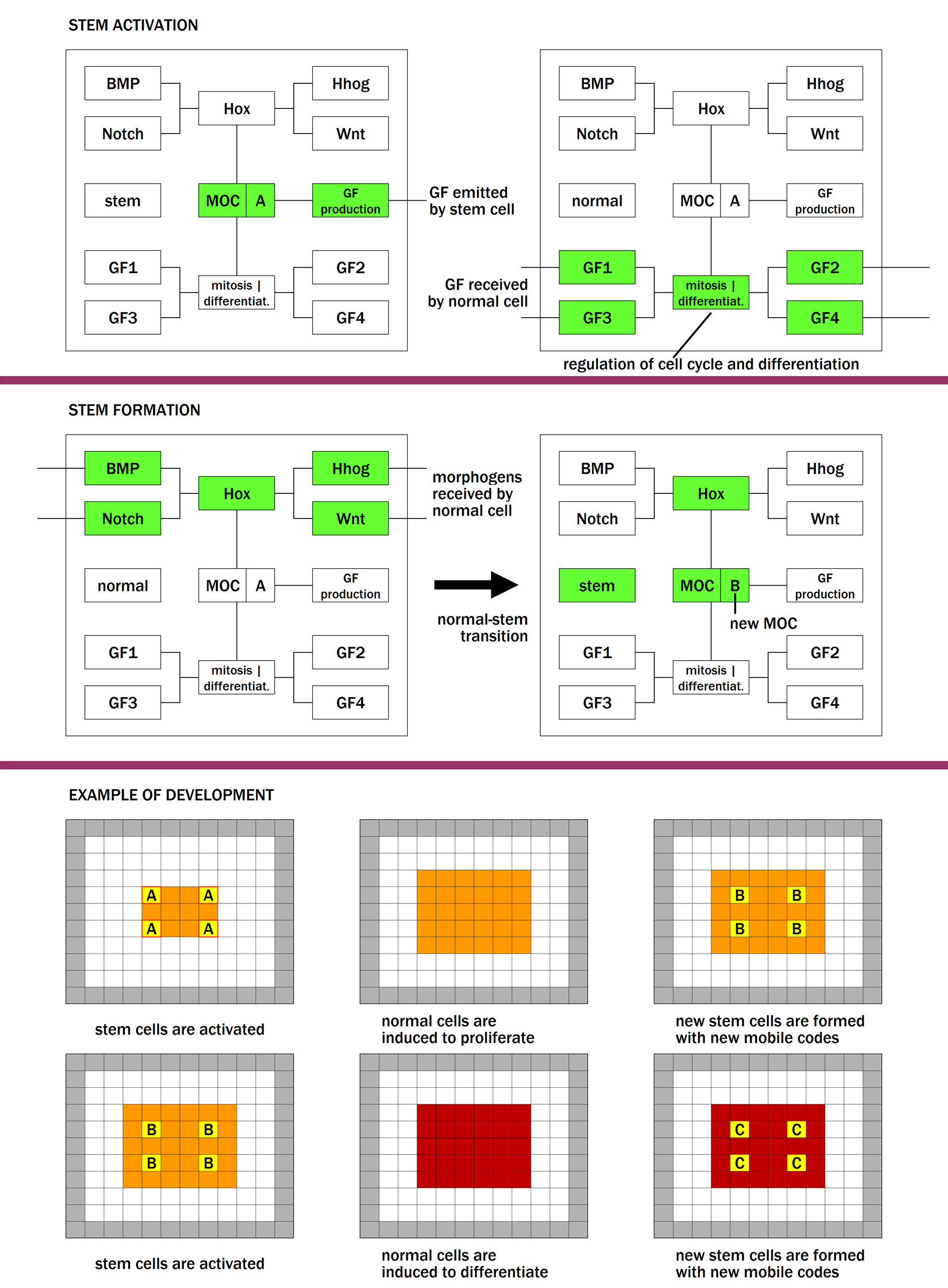}}}
\caption{\textbf{Upper panel:} stem activation. On the left side, a stem cell is activated and releases growth factors into the external environment. On the right side, the growth factors are received by neighbouring normal cells, which are induced to proliferate or to differentiate. \textbf{Middle panel:} stem formation. On the left, a normal cell receives morphogens from the external environment. Their interplay determines the creation of a new Hox protein, which contributes to determine the establishment of a new mobile code (on the right). \textbf{Lower panel:} example of tissue development composed of two rounds of the stem activation-stem formation cycle.}
\label{framen}
\end{center} \end{figure}

This abstract model of development can be refined by linking the cellular behaviour with modules which reproduce specific intracellular mechanisms, such as Hox genes, morphogen- and growth factor-related pathways. A scheme of the interplay of these factors and their relation to the phenomena of stem activation and stem formation is reported in Fig.~\ref{framen}. The mobile code of a stem cell can trigger the activation of the \textit{growth factor production module}, dedicated to the production of growth factors (Fig.~\ref{framen}, upper panel, left part). This module contains the macro genes described above. The growth factors exit the cell and are received by a neighbouring normal cell (Fig.~\ref{framen}, upper panel, right part), and are processed by dedicated \textit{growth factor (GF) modules}. As a result, the cell enters the cell cycle and/or activates other genes needed for differentiation.  

\begin{figure}[p] \begin{center}
{\fboxrule=0.0mm\fboxsep=0mm\fbox{\includegraphics[width=16.50cm]{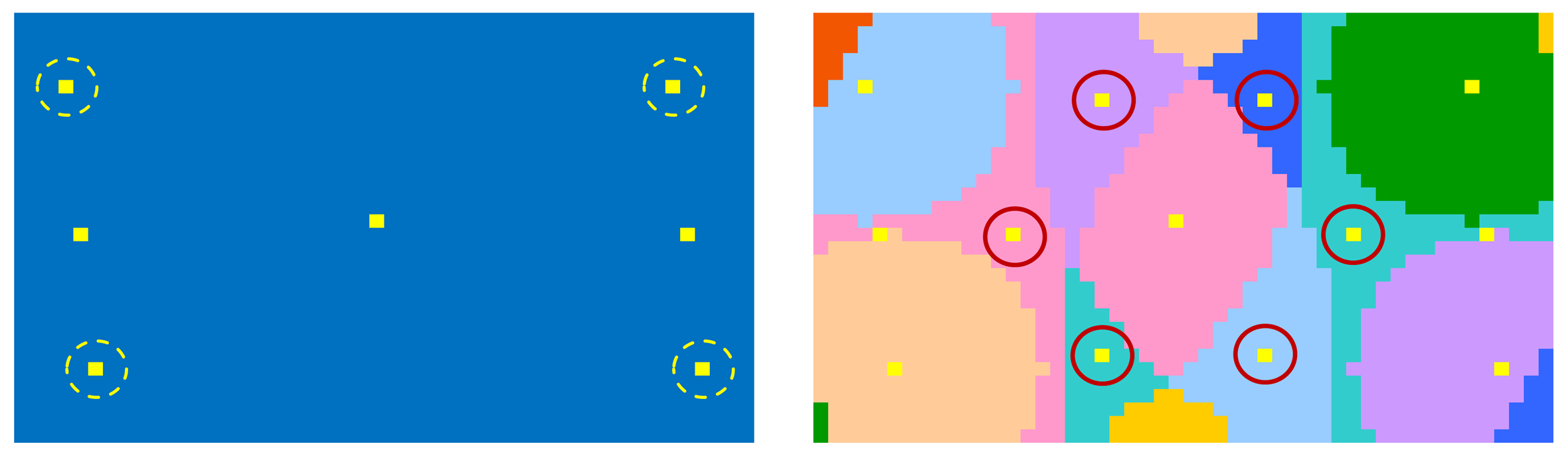}}}
\caption{Morphogens in ET. The panel on the left shows a hypothetical distribution of stem cells. Some of these stem cells (dashed circles) emit morphogens, of a finite number of types, which diffuse across the space. On the right, the different combinations of concentration values of the morphogens partition the space into regions (marked each with a different colour). In each such region a new stem cell is created from a normal cell (red circles).}
\label{signals}
\end{center} \end{figure}

In ET the process of normal-stem conversion is achieved through the presence of simulated morphogens, with the following mechanism. Some stem cells (Fig.~\ref{signals}) become sources of specific morphogens (belonging to a number of distinct types). These signals partition the space into regions, each characterised by a distinct combination of concentration values. For example, all cells in a given region ``sense'' one morphogen with concentration 7, another morphogen with concentration 4, etc. (concentration values are integer numbers which represent categories or ranges of actual concentration values). In each region a normal cell is selected and turned into a stem cell.

\begin{figure}[p] \begin{center} \hspace*{-0.50cm}
{\includegraphics[width=17.50cm]{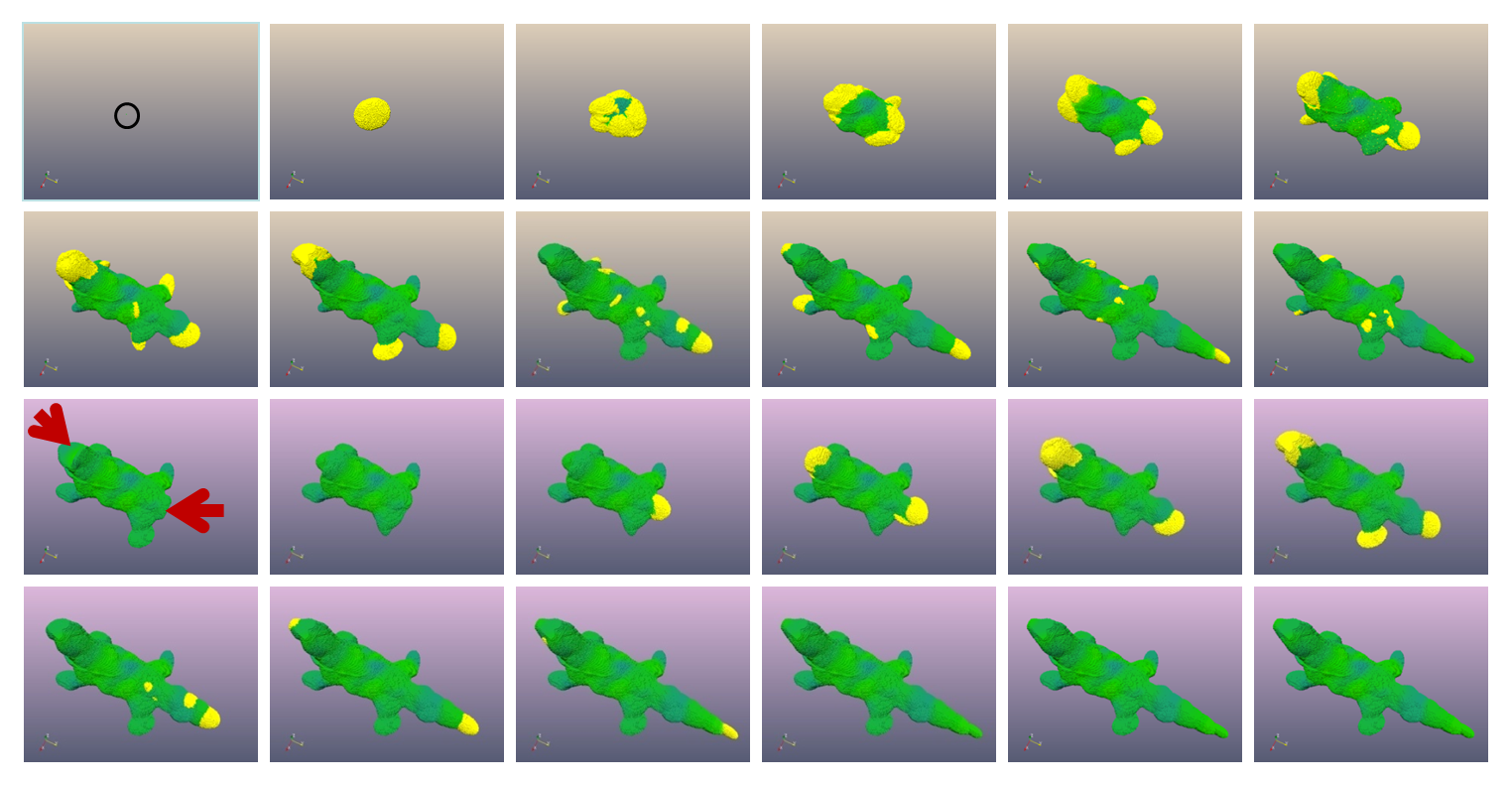}}
\caption{Self-generation and self-regeneration of a lizard-shaped structure consisting of 500000 cells in ET. Development unfolds from a single cell (circle) in 12 developmental stages (frames 1-12). After stage 12, the head and the tail are cut off (arrows). After the ``debris'' is removed, the structure is completely regenerated.}
\label{regsimul}
\end{center} \end{figure}

\textit{Morphogen processing (MP) modules} are dedicated to the processing of morphogens. For simplicity we represented only four MP modules, named as the corresponding biological pathways (BMP, Hedgehog, Notch and Wnt). The process of new stem formation starts when a normal cell receives morphogens from the external environment (Fig.~\ref{framen}, middle panel, left part). These signals are processed by the MP modules and by the \textit{Hox module}. The outcome is represented by one or more Hox proteins, which are added to the repertoire of master transcription factors already present in the cell, and contribute to determine the new mobile code (Fig.~\ref{framen}, middle panel, right part). The lower panel shows an example of tissue development, obtained with two rounds of the stem activation-stem formation cycle described.

\colorbox{colmb}{According to}this model, in biological cells each MP pathway senses the concentration of a specific morphogen, and sends a signal into the nucleus. Here the interplay of signals conveyed by all pathways results in the transcription of a different Hox gene. For example, if the combination of concentrations is [2130] (i.e. one morphogen is received with concentration 2, another morphogen with concentration 1, etc.) Hox gene X is transcribed. If the combination of concentrations is [3110], Hox gene Y is transcribed. The particular combinations of morphogen concentrations necessary for mobile code generation may correspond to the set of signals required to maintain stem cells in niches \cite{Song04}.

\colorbox{colmb}{ET foresees}that each organ or anatomical structure of the body is assembled during embryogensis thanks to the cooperation of a set of morphogens and a set of growth factors, which are organ-specific. Table~\ref{organs} gives an overview of a hypothetical involvement of four morphogens (a, b, c, d) and four growth factors (p, q, r, s) in four organs. For each organ three morphogens and three growth factors are hypothesised to be necessary to achieve development and differentiation of all organ components. We hypothesise that this is true also in biological systems. 

\begin{table}[ht]
\vskip 0.25cm
\center{
\begin{tabular}{|l      | C{2cm}  | C{2cm}  | C{2cm}  | C{2cm}  |} \hline
organ                   & organ 1 & organ 2 & organ 3 & organ 4 \\ \hline
morphogens involved     & a,b,c   & a,b,d   & a,c,d   & b,c,d   \\ \hline
growth factors involved & p,q,r   & p,q,s   & p,r,s   & q,r,s   \\ \hline
\end{tabular}}
\vskip 0.25cm
\caption{Hypothetical involvement of morphogens and growth factors in the embryonic development of four organs.}
\label{organs}
\end{table}

The proposed framework allows also to model cellular regeneration. Our concept of regeneration is based on the idea to recreate exactly the same conditions as during development. Stem cells, after their activation during development, are kept in a deactivated state (so that the associated event is not triggered again): these correspond in nature to adult stem cells. Once a damage is detected in a given body part, the stem cells activated to produce that body part during development are reactivated during regeneration, leading to the same sequence of events, and restoration of the damaged part (Fig.~\ref{regsimul}).



\section{Evo-devo and transposable elements}
\label{sec:evo-devo}

\begin{figure}[t] \begin{center} \hspace*{-0.50cm}
{\fboxrule=0.0mm\fboxsep=0mm\fbox{\includegraphics[width=17.50cm]{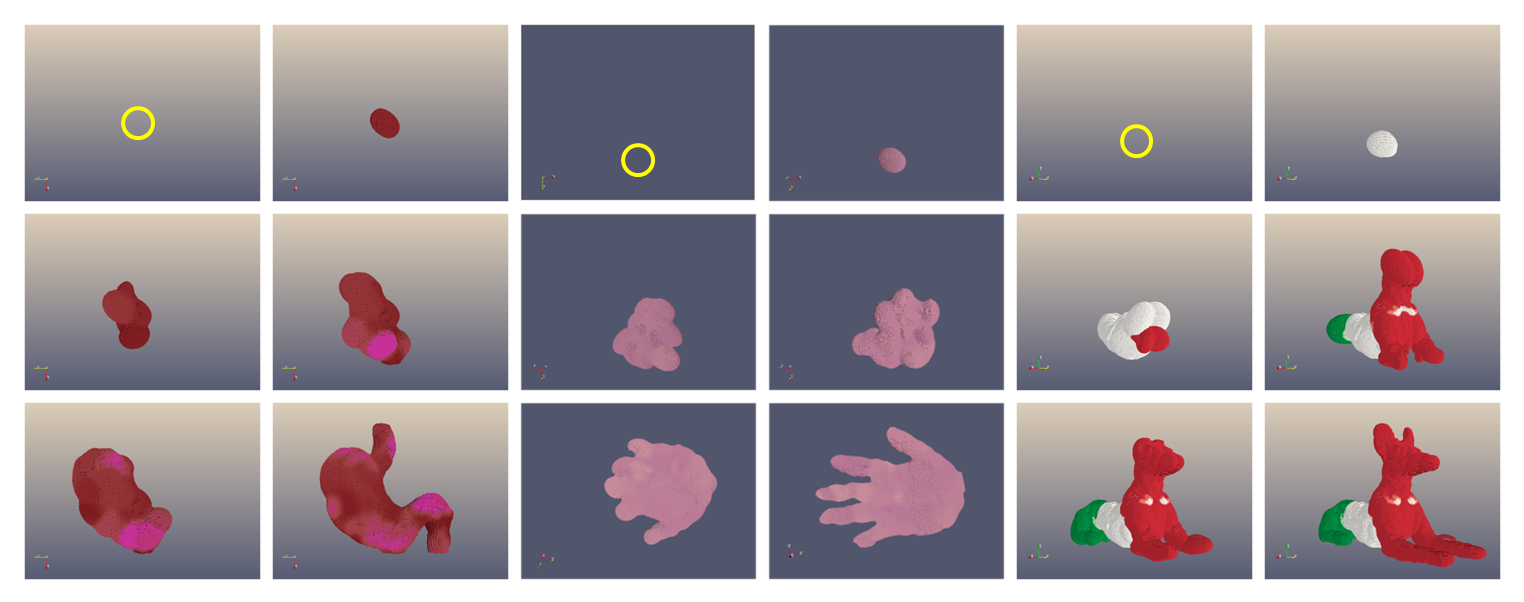}}}
\caption{Examples of structures generated using ET. The panels show some stages of development from the single cell stage (yellow circles) for three structures. The structures, composed of millions of cells, represent a stomach, a hand, and a dog patterned with the colours of the Italian flag. This last one is a more complex version of the classical Wolpert's French frag model.}
\label{devsimul}
\end{center} \end{figure}

The model of development described can be coupled with a \textit{genetic algorithm}, able to simulate Darwinian evolution. The genetic algorithm evolves a population of individuals (each encoded in an artificial genome) for a number of generations. At each generation, all individuals develop independently from the zygote stage to the final phenotype, whose proximity to a predefined target is employed as a fitness measure. This operation is repeated for all individuals, so that eventually each individual is assigned a fitness value. Based on this value the genomes of the individuals are selected and randomly mutated, to produce a new population. This cycle is repeated until a satisfactory level of fitness is reached. The coupling of the model of development and the genetic algorithm gives origin to an evo-devo process, which was proved able through computer simulations to ``devo-evolve'' 3-dimensional structures of unprecedented complexity (see examples in Fig.~\ref{devsimul}). 

\begin{figure}[t] \begin{center} \hspace*{-0.50cm}
{\fboxrule=0.0mm\fboxsep=0mm\fbox{\includegraphics[width=17.50cm]{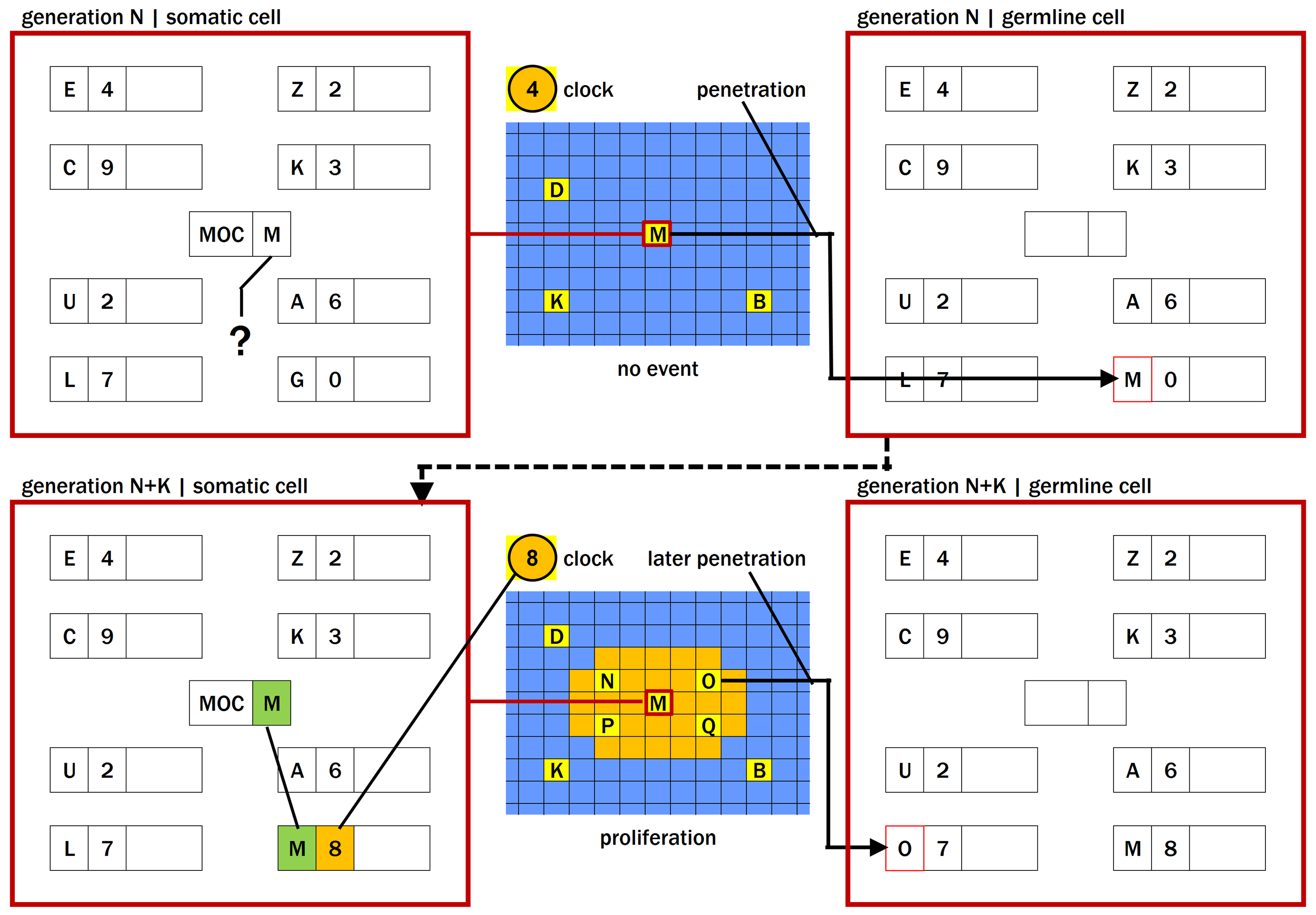}}}
\caption{Germline penetration of new regulatory sets in ET. In the upper part, in a stem cell the genome is not able to ``respond'' to the mobile code M with a corresponding event. A new regulatory set able to respond is created in the stem cell. The new regulatory set leaves the stem cell and reaches the germline, where it is incorporated in the genome. In a subsequent generation, when a cell has mobile code M, its genome (inherited from the penetrated germline) can respond with a specific macro gene. The result is a proliferation event.}
\label{noevent}
\end{center} \end{figure}

In ET most stem cells produced during development do not orchestrate any events (are inactive), because in the genome there is no macro gene whose regulatory set can match their mobile code (upper left part of Fig.~\ref{noevent}). The probability that a suitable regulatory set (which is encoded as a long sequence of numbers) emerges in the genome simply through mutations and recombinations is very low. A countermeasure consists in ``suggesting'' to the genetic algorithm to put in the genome some regulatory sets which are guaranteed to match. This idea is implemented in a procedure called \textit{germline penetration}, which builds regulatory sets able to match mobile codes generated during the development of the structure. These regulatory sets are then associated to macro genes contained in a special copy of the genome called ``germline'' genome (upper right part of Fig.~\ref{noevent}) which, after reproduction, is destined to become the (``somatic'') genome of the individuals of the next generation. 

Once evolution is provided with a ``good'' regulatory set, i.e. one guaranteed to match an existing mobile code, it has to optimise the coding part of the macro gene, a process that can take several generations. When the optimisation of the coding part is completed, the new macro gene can be activated and carry out a developmental event (lower part of Fig.~\ref{noevent}). New mobile codes are generated as a result of the new event: regulatory sets able to match such new mobile codes are again transferred to the germline genome, and the whole cycle repeats itself. The penetrated macro genes are initially set as inactive; otherwise, the gene would become active with a non-optimised coding part, causing a disruption in development (and an abrupt decrease in the individual's fitness). Their activation is obtained through a subsequent genomic mutation of the switch. Therefore, according to ET, evolution is essentially Lamarckian for regulatory sequences and Darwinian for coding sequences.
 
Germline penetration is essential in ET for the evolution of complex shapes: in our \emph{in silico} experiments, if germline penetration is disabled, the evolutionary process practically grinds to a halt. The central role played by germline penetration in our model lead us to hypothesise the existence of a similar process also in biological organisms. For the implementation of a biological germline penetration, we need genetic elements able to build new regulatory sequences in somatic cells, and susceptible of being transferred to germline cells.

\colorbox{colmb}{Transposable elements,}or transposons \cite{Mcclintock50}, are DNA sequences that can move around to different positions in the genome. During this process, they can cause mutations, chromosomal rearrangements and lead to an increase in genome size. Despite representing a large genomic fraction (30-40\% in mammals), transposons have no clear biological function, and have therefore been labelled as ``junk DNA''. Transposable elements are active during development, and may bring diversity in somatic cells having the same genome \cite{Collier07}. A transposonal activity was also observed in germline cells, with possible evolutionary implications. ``Waves'' of diffusion of transposable elements in a genome appear temporally associated with major evolutionary changes to the species \cite{Oliver10}, suggesting a causal link between the two events. \nr
 
\colorbox{colmb}{Based on}these considerations, we can imagine a mechanism for the implementation of a biological germline penetration. In a stem cell unable to produce a developmental event (because the set of master transcription factors present in the cell cannot bind to corresponding regulatory sequences), transposable elements would become mobilised, and start ``jumping'' around the genome. This process could lead to the creation of many new regulatory sequences, hugely increasing the chances that a lucky combination of such sequences be able to match the existing transcription factors. We hypothesise that, when a match occurs, the transposable elements which helped to build the new sequences leave the cell and, through the bloodstream, make their way to germline cells. Here they insert themselves into the genome, thus allowing the delivery of a successful innovation to the next generation.
 
The mechanism described has interesting evolutionary implications. In ET, whenever a stem cell proliferates, a wave of new stem cells, with new mobile codes, is created in the body of the (new) species. The action of germline penetration translates this wave of new mobile codes in somatic cells into a corresponding wave of new regulatory sets spreading in the germline genome and subsequently in the somatic genome of future generations. Such events during artificial evolution coincide with moments in which major changes occur to the evolving species, causing new body parts or features to appear. 

\colorbox{colmb}{In biological}terms, this means that the spreading in the genome of waves of new sets of transposable elements in the course of evolution corresponds to moments in which new branches (new species) are generated in the ``tree of life''. Such predictions, made with our model on purely theoretical grounds, appear to be confirmed by experimental evidence \cite{Oliver10}. These observations suggest that the spreading of new transposon families in the genome and the occurrence of major changes in the relevant lineage are simultaneous events. This seems to hint that the colonisation of the genome by the transposons is the driving force behind the change in the lineage. Our interpretation of this phenomenon is different. In fact, according to ET, the spread of new transposon families in the genome is an event which comes immediately (in evolutionary terms) {\em after} the change, not before. The confirmation of this prediction, made possible by techniques to estimate the age of DNA sequences more sensitive than those currently employed, would represent a clear indication in favour of this model.
  
In conclusion, our model suggests a biological role for transposable elements which, as we have seen, account for up to 40\% of the genomic content in mammals. We argue that the presence of trasposable elements is not an artifact, but an essential biological feature. In the ET framework, for any individual, the set of mobile codes generated during development can be divided into i) mobile codes that activate a macro gene during development and ii) mobile codes that do not activate any macro gene during development. Indeed, the ET machine cannot avoid generating inactive mobile codes. To prevent this effect we should reduce the density of stem cells, but this would also reduce the effectiveness of the morphogenetic process. Therefore, the presence of a certain amount of inactive mobile codes is unavoidable. On this material intervenes germline penetration, converting inactive mobile codes into a corresponding number of inactive regulatory sets in the genome (the penetrated regulatory sets are inactive because the associated switches are initially set to OFF). Therefore, the presence of inactive information in both the set of mobile codes and the genome is a fact which is inescapably connected to the core of the ET machine, a requirement essential to its \emph{evolvability}.



\section{Ageing}
\label{sec:ageing}
 
\colorbox{colmb}{Ageing is}a universal phenomenon, for which many theories exist. \emph{Stochastic theories} (e.g. the ``free-radical theory'' \cite{Gerschman54}) blame damage induced by environmental factors, and accumulating over time, as the cause of ageing. According to \emph{programmed theories} (e.g. the ``ageing-clock theory'' \cite{Dilman92}), ageing is driven by genetic instructions, which can to a certain extent be modulated by environmental conditions. \emph{Evolutionary theories} (e.g. the ``disposable soma theory'' \cite{Kirkwood77}) see evolution as the main force shaping the ageing profile of different species.
 
\colorbox{colmb}{A measure}that was proved effective in reducing the rate of ageing in many species is dietary restriction \cite{Masoro07}. Many genes have been shown to influence the life span in some species; most of these genes belong to the insulin / IGF1-like signalling pathway, to the class of sirtuins, and to the target of rapamycin (TOR) pathway \cite{Kenyon10}. This last pathway is thought to mediate the effects of dietary restriction at the cellular level. Overall, it is safe to say that both genetic and environmental factors concur to cause ageing, but the balance between these factors appears to change in the course of life: the first seem to prevail in the young, the second become dominant in the old. We will now propose a possible explanation for this phenomenon.
  
In ET, for a given individual development unfolds in N developmental stages; at the end of it the individual's fitness is evaluated, and used for reproduction in the genetic algorithm. The moment of fitness evaluation, that in nature may correspond to the moment of reproduction, in most of our experiments has coincided with the end of the simulation. On the other hand, we can imagine to let the global clock tick on and see what happens in the period after fitness evaluation. The distinction between the periods before and after fitness evaluation can be thought to correspond to the biological periods of development (say, until 25 years of age in humans) and ageing (from 25 years of age onwards). 

\begin{figure}[p] \begin{center}
{\fboxrule=0.0mm\fboxsep=0mm\fbox{\includegraphics[height=06.00cm]{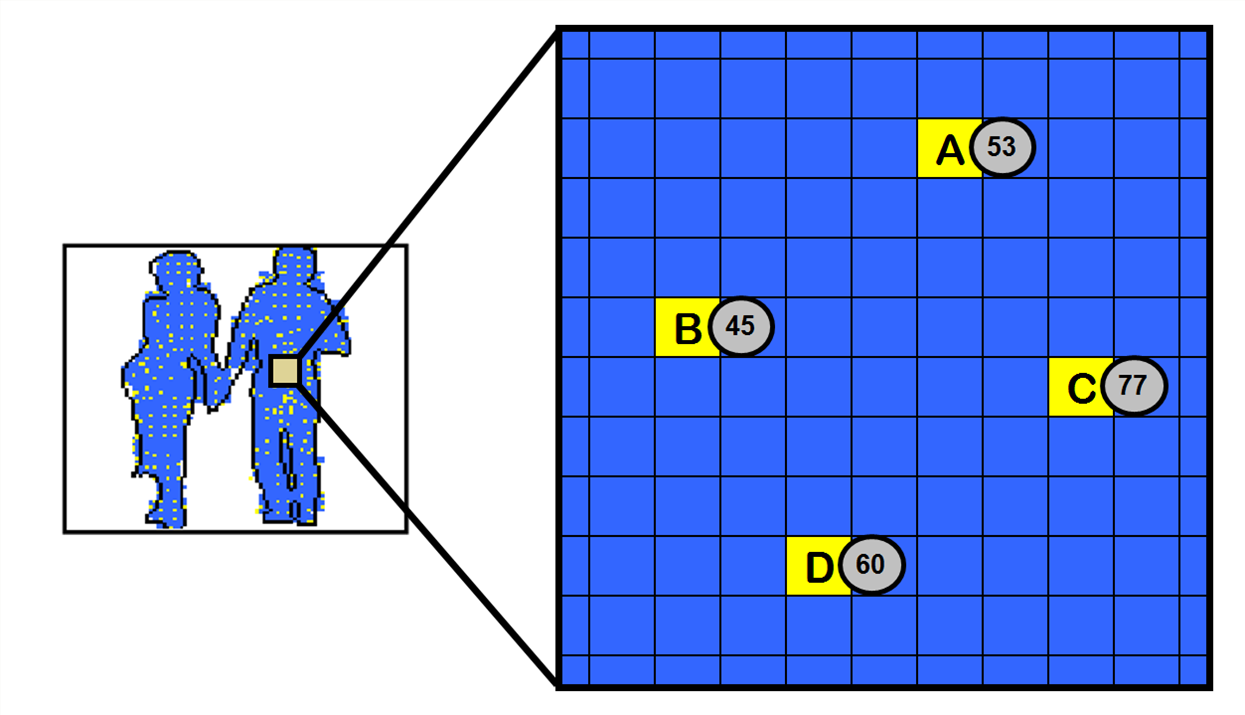}}}
\caption{Stem cells present in ET artificial bodies at the end of development. Some of these stem cells have been activated during development, some (the vast majority) have not. The timer value of the gene which is going to be activated in each stem cell is indicated in the circles (numbers refer to age in humans expressed in years).}
\label{tmcells}
\end{center} \end{figure}

\begin{figure}[p] \begin{center}
{\fboxrule=0.0mm\fboxsep=0mm\fbox{\includegraphics[width=15.00cm]{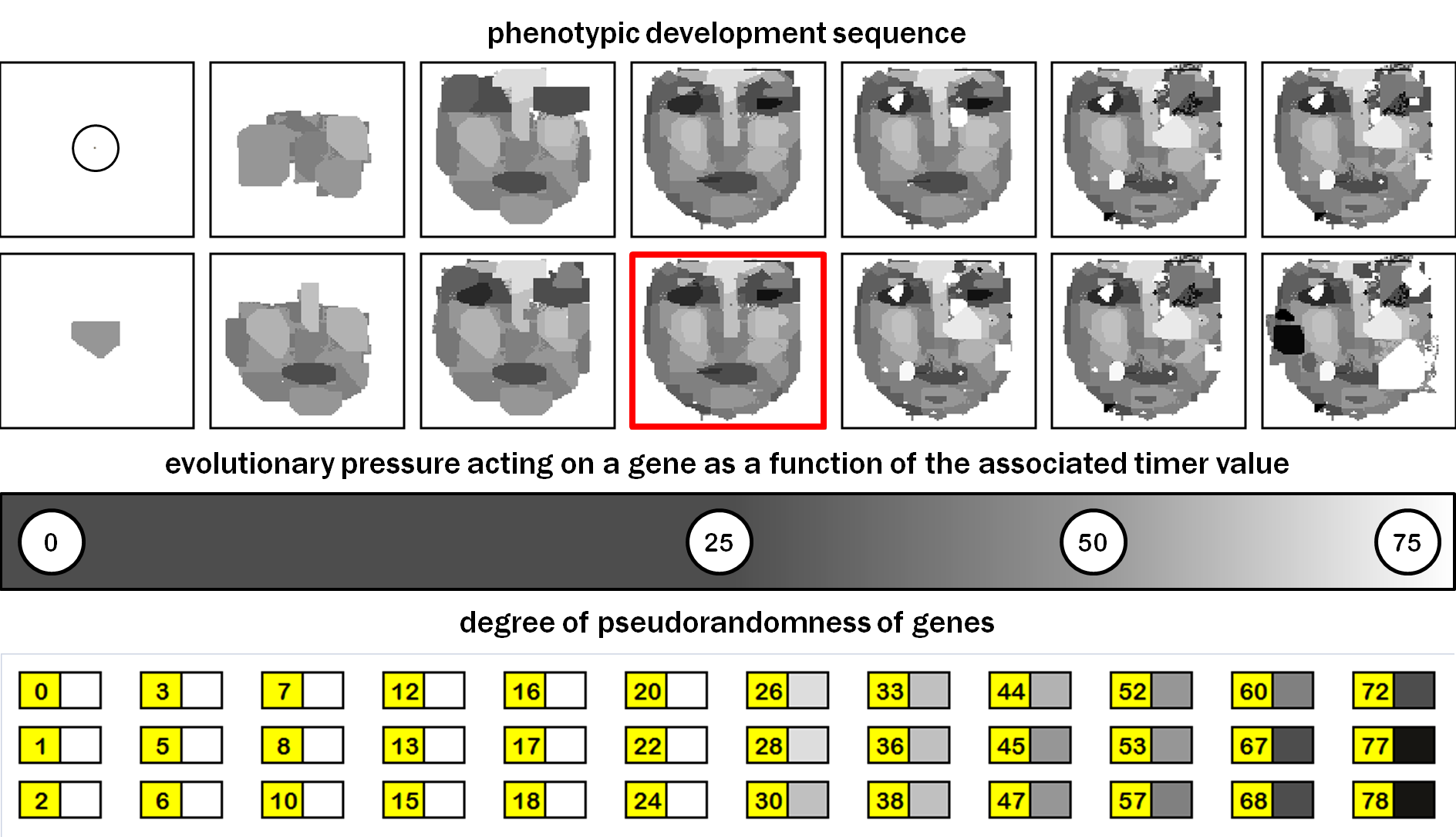}}}
\caption{In the upper panel: example of ageing in ET for a ``face''. On the left the period of development: the structure grows from a single cell to the mature phenotype at age 25 (red frame), when fitness is evaluated. On the right the period of ageing: the quality of the structure deteriorates steadily under the action of non-optimised genes. In the middle panel: the evolutionary pressure acting upon a gene (represented by the darkness level) is an inversely proportional function of the timer value. Genes with timer values lower than 25 years are subject to a high evolutionary pressure, genes with timer values greater than 25 years are subject to a steadily decreasing pressure, until their effects become pseudorandom. In the lower panel: the level of pseudorandomness (represented by the darkness level) of a gene coding part increases with the gene timer value.}
\label{ageface}
\end{center} \end{figure}

At the end of an individual's development, many stem cells are present in the body of the individual. Some of these stem cells have been activated during development (and shaped the individual's body), some (the vast majority in our simulations) have not (Fig.~\ref{tmcells}). These stem cells may contain genes bound to be activated in the ageing period, after the moment of fitness evaluation (the moment of activation is determined by their timer value, $\ge$ 25) when, by definition, they are not affecting the fitness value. For this reason, their coding parts have not been optimised by evolution and the associated events will tend to have a ``pseudorandom'' character, i.e. they will look random, even though they are not (they are encoded in the genome and, as such, entirely under genetic control and deterministic). 

The consequence of the ``pseudorandomness'' of these events is that their effects on the overall individual's health are more likely to be detrimental than beneficial. Here the term ``health'' is used to indicate the individual ``physical'' condition, while the (reproductive) ``fitness'' corresponds to the chance that the individual's genes will survive in the future genetic pool of the population. As a result, in the ageing period, the individual's health will tend to progressively decrease over time, under the action of such pseudorandom events. Therefore, ET sees ageing as a \emph{continuation of development}, driven by non-optimised genes activated in specific stem cells after the age of reproduction. Our postulate is that these condiderations apply also to biological ageing.
 
The hypothesis, according to which the evolutionary pressure acting on a gene is null if it is activated after reproduction, is rather simplistic. In nature, an individual's reproductive fitness depends also on events manifesting themselves after reproduction, as also these affect the survival chances of its progeny. More realistically, the effect of an event on the fitness (and hence the evolutionary pressure acting upon the corresponding gene) will tend to decrease as the age of its manifestation increases, rather than vanishing immediately after reproduction. Figure~\ref{ageface} shows a simulation of ageing for a ``face'', which incorporates this last observation.
 
\colorbox{colmb}{The theory}proposed provides a synthesis of all three categories of ageing theories. It has a solid evolutionary background, and explains why the experimental evidence has a deterministic connotation at young age and turns stochastic with age progression. It has also the advantage to interpret development and ageing with a common theoretical framework. In biology, besides proliferation and apoptosis, we can envision the existence of other types of events. Another ``biological change event'' could induce epigenetic changes in the gene regulatory networks of affected cells. This, in turn, would change the biochemistry of the relevant tissue and the function of the organ.

\colorbox{colmb}{Another merit}of our proposal is the involvement of stem cells in the picture. By analogy with ``cancer stem cells'', we may call the stem cells involved in the ageing process ``ageing stem cells''. There is empirical evidence of an age-related decline in the functionality of adult stem cells \cite{Liu11}. Our hypothesis is that this functional decline is determined by epigenetic changes induced in stem cells by biological change events, bound to occur at precise moments. Such biological change events may occur both in stem cells activated during embryonic development, and reactivated in the post-developmental period for tissue repair, or in stem cells activated in the post-developmental period for the first time. Both types of events are hypothesised to contribute to the ageing process.
 
\colorbox{colmb}{The view}of ageing as the progressive accumulation of pseudorandom events provides a new interpretation for diseases typically associated to the old age (Alzheimer disease, type II diabetes, etc.), which seem to be caused by the malfunctioning of specific genes, but whose onset in humans typically occurs from the 5th decade of life onwards. Our hypothesis is that the difference between these diseases and the effects of ``normal'', ``healthy'', ageing is more quantitative than qualitative. We think that both ageing and ageing-associated diseases are driven by biological change events caused by genes set to be activated in the old age: the phenotipic manifestations of normal ageing are only milder, more benign than those associated to such diseases. This interpretation explains why the temporal patterns of ageing and age-related diseases are coincident, and finds a strong confirmation in the fact that measures known to delay ageing, such as caloric restriction, also postpone the onset of diseases \cite{Colman09}.
 
\colorbox{colmb}{As we}mentioned, a slow-down in the pace of ageing can be obtained through dietary restriction. One possible explanation of this evidence is that dietary restriction directly affects the functioning of a biological equivalent of the global clock, which becomes slower. As a result, all genes bound to be activated in the ageing period are delayed, all by the same amount. The genetic pathways known to influence life span may be part of the molecular device which transduces the clock value (perhaps implemented as a diffusible molecule) from the external cellular environment to the nucleus of cells.   
   


\section{Cancer}
\label{sec:cancer}
 
\colorbox{colmb}{Cancer is}a class of diseases involving unregulated cell growth. According to the ``standard theory'', also referred to as the ``multi-hit'' hypothesis \cite{Knudson71}, carcinogenesis is a multi-step process that can take place in any cell, driven by multiple damages (``hits'') to genes that normally regulate cell proliferation and cell death . An alternative theory considers cancer stem cells responsible for tumour growth \cite{Bonnet97}. Cancer stem cells share with normal stem cells the ability to self-renew and differentiate into multiple cell types: the presence in tumours of cells of different types and/or having different degrees of differentiation is a well documented phenomenon, coherent with the cancer stem cell theory.

\colorbox{colmb}{In recent}years, new cancer stem cell models have been proposed, e.g., the ``complex system model'' \cite{Cruz12}, or the ``stemness phenotype model'' \cite{Laks10}. These models share the ideas that all cells in a tumour are potentially tumorigenic, and that stemness is a dynamic property. The stem-non stem conversion would occur in both directions, triggered by genetic and epigenetic factors, and influenced by the cellular microenvironment.

\colorbox{colmb}{A critical}point for all cancer theories is the impossibility to reconduct tumour formation to a common set of gene mutations. A few cancer-related genes, such as p53, do seem to be mutated in the majority of tumours, but many other cancer genes are changed in only a small fraction of cancer types, a minority of patients, or a subset of cells within a tumour. Although the effort to reconduct tumour formation to subsets of mutated genes has been unsuccessful, it is nonetheless undeniable that correlations between tumours and patterns of mutations exist, i.e. individual genes are mutated in percentages that are tumour-specific \cite{Yeang08}.

\begin{figure}[p] \begin{center} \hspace*{-0.5cm}
{\fboxrule=0.0mm\fboxsep=0mm\fbox{\includegraphics[width=16.00cm]{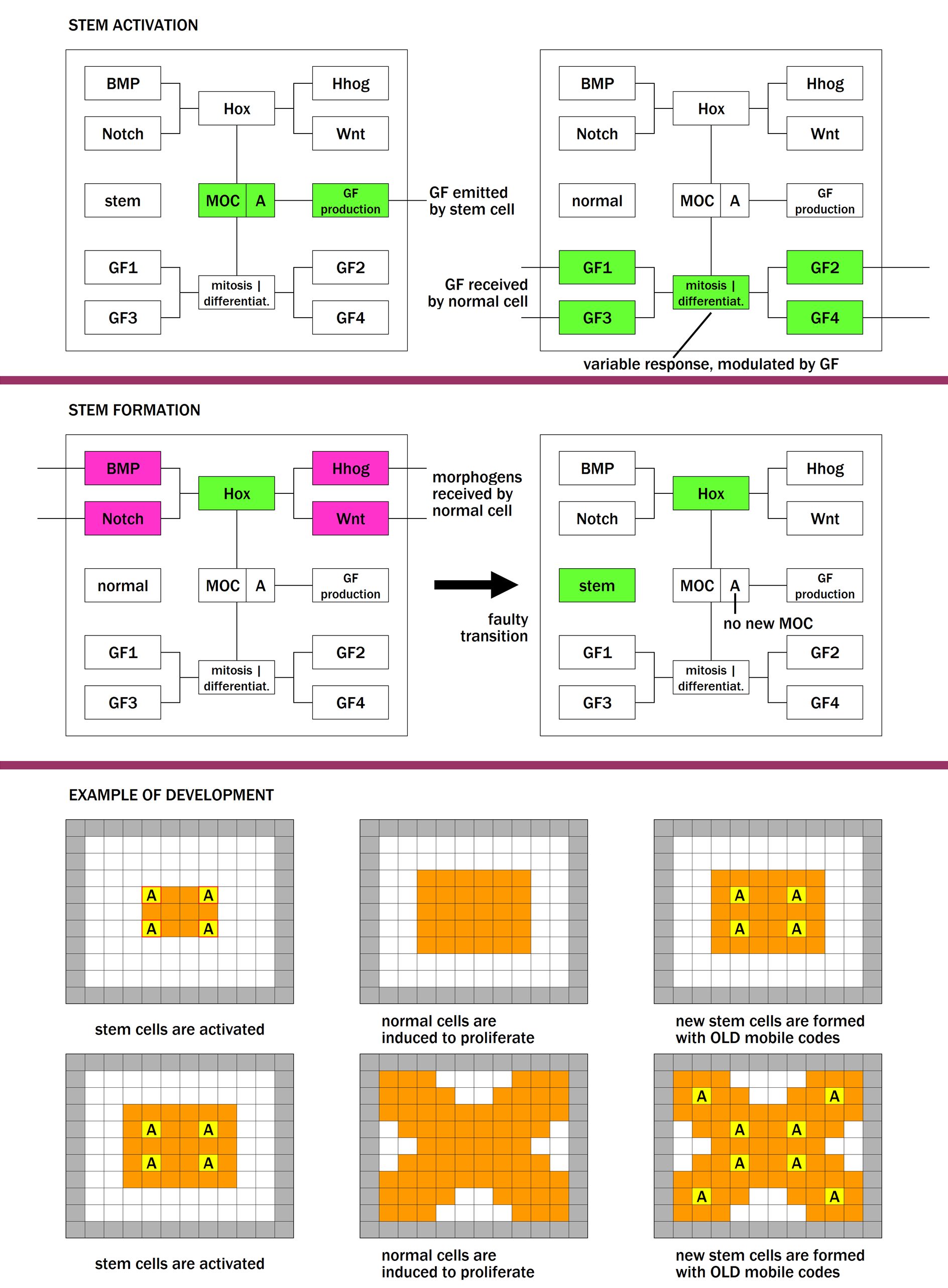}}}
\caption{When one or more MP module are damaged, the mobile code generation device is impaired (\textbf{middle panel}). As a result, mobile codes B necessary to issue the differentiation commands cannot be created. Instead, mobile codes A are produced, which induce cells to keep proliferating: the final outcome is an endless chain of proliferations of immature cells (\textbf{lower panel}). This, despite the fact that cells preserve a normal response to proliferation and differentiation signals (\textbf{upper panel}).}
\label{framekm}
\end{center} \end{figure}

\begin{figure}[p] \begin{center} \hspace*{-0.5cm}
{\fboxrule=0.0mm\fboxsep=0mm\fbox{\includegraphics[width=16.00cm]{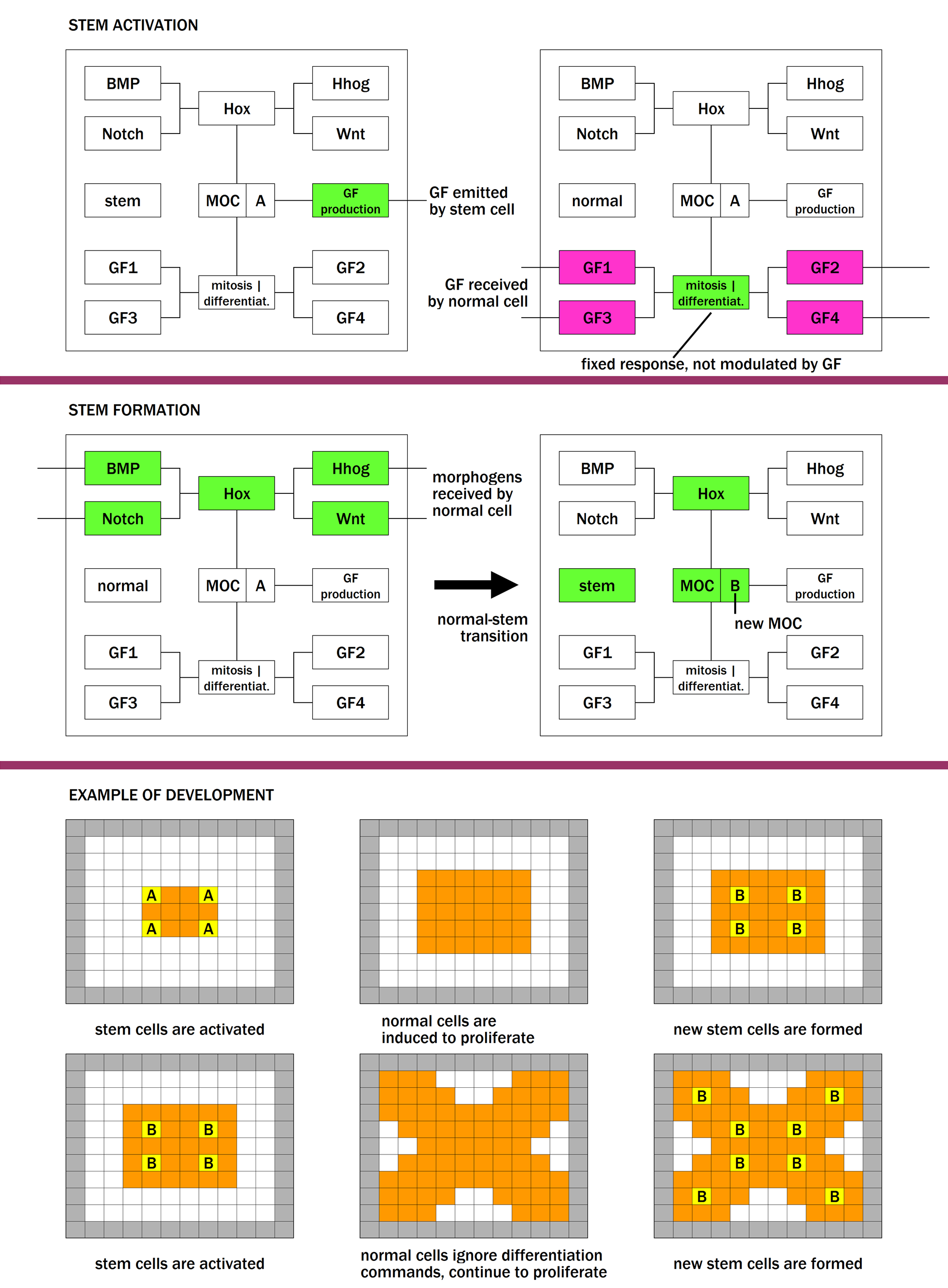}}}
\caption{When the damage hits one or more GF modules, normal cells ignore commands coming from stem cells (i.e. growth factors) and continue to proliferate under circumstances in which they would normally exit the cell cycle and differentiate (\textbf{upper panel}). The final outcome is an endless chain of proliferations of immature cells (\textbf{lower panel}). This, despite the fact that the normal-stem mechanism is intact and the commands received by normal cells are correct (\textbf{middle panel}).}
\label{framekf}
\end{center} \end{figure}

\colorbox{colmb}{As we}have seen in section 2, ET foresees that each organ or anatomical structure of the body is assembled, during embryogensis and during tissue regeneration, through the action of MP modules and GF modules. If one or more such modules are damaged, the assembly mechanism may not function properly.  

\colorbox{colmb}{If one}or more MP modules are damaged, the mobile code generation device is impaired (Fig.~\ref{framekm}). As a result, mobile codes B necessary to issue the differentiation commands cannot be created. Instead, mobile codes A are produced, which induce cells to keep proliferating. The final outcome is an endless chain of proliferations of immature cells. A damage to MP modules in our model may correspond in biology to mutations affecting genes involved in MP pathways. Interestingly, genes belonging to such pathways are frequently mutated or aberrantly activated in cancer \cite{Karamboulas13}. 

\colorbox{colmb}{In our}interpretation, a fully working differentiation machinery requires the functioning of \emph{all} MP pathways involved in the generation of the organ. On the other hand, a reduced form of differentiation can be obtained in many different ways, corresponding to all possible different subsets of non-functional MP pathways. If a sufficient number of genes in one or more MP pathways are rendered non-functional (through mutations), differentiation fails, and cancer ensues. 

\colorbox{colmb}{Another dangerous}situation occurs when the damage hits one or more GF modules (Fig.~\ref{framekf}). In this case, normal cells ignore commands coming from stem cells (i.e. growth factors) and continue to proliferate under circumstances in which they would normally exit the cell cycle and differentiate. Even in this case we can hypothesise that, while a fully working cellular machinery requires the functioning of all GF pathways involved, a reduced form of functioning can be obtained in many different ways, corresponding to all possible different subsets of non-functional GF pathways. 

\colorbox{colmb}{Table~\ref{cancers}}refers to the same morphogens and growth factors listed in Table~\ref{organs}. The row labelled ``combinations of damaged MP pathways'' lists all possible combinations of damaged MP pathways which lead to failed differentiation (and hence tumour formation) in the relevant organ. Different combinations of damaged pathways may correspond to tumours of different grades. Considering that a damage in each MP pathway can be caused by mutations to a number of genes, this scheme can help to explain the heterogeneity of the mutational patterns observed, and the presence in tumours of cells having different degrees of differentiation and/or of different cell types. Analogous considerations can be applied to GF pathways.

\begin{table}[ht]
\vskip 0.25cm
\center{
\begin{tabular}{|l      | C{2cm}    | C{2cm}    | C{2cm}    | C{2cm}    |} \hline
organ                   & organ 1   & organ 2   & organ 3   & organ 4   \\ \hline
morphogens involved     & a,b,c     & a,b,d     & a,c,d     & b,c,d     \\ \hline
combinations of         & \{a\}     & \{a\}     & \{a\}     & \{b\}     \\
damaged MP pathways     & \{b\}     & \{b\}     & \{c\}     & \{c\}     \\
                        & \{c\}     & \{d\}     & \{d\}     & \{d\}     \\
                        & \{a,b\}   & \{a,b\}   & \{a,c\}   & \{b,c\}   \\
                        & \{a,c\}   & \{a,d\}   & \{a,d\}   & \{b,d\}   \\
                        & \{b,c\}   & \{b,d\}   & \{c,d\}   & \{c,d\}   \\
                        & \{a,b,c\} & \{a,b,d\} & \{a,c,d\} & \{b,c,d\} \\ \hline
growth factors involved & p,q,r     & p,q,s     & p,r,s     & q,r,s     \\ \hline
combinations of         & \{p\}     & \{p\}     & \{p\}     & \{q\}     \\
damaged GF pathways     & \{q\}     & \{q\}     & \{r\}     & \{r\}     \\
                        & \{r\}     & \{s\}     & \{s\}     & \{s\}     \\
                        & \{p,q\}   & \{p,q\}   & \{p,r\}   & \{q,r\}   \\
                        & \{p,r\}   & \{p,s\}   & \{p,s\}   & \{q,s\}   \\
                        & \{q,r\}   & \{q,s\}   & \{r,s\}   & \{r,s\}   \\
                        & \{p,q,r\} & \{p,q,s\} & \{p,r,s\} & \{q,r,s\} \\ \hline
\end{tabular}}
\vskip 0.25cm
\caption{Hypothetical involvement of four MP pathways and four GF pathways in as four cancer types.}
\label{cancers}
\end{table}

\colorbox{colmb}{Based on}this scheme, ``driver'' mutations found in cancer samples can be divided in three broad categories: mutations to MP pathways, mutations to GF pathways, and mutations to other genes, which confer to the trasformed cell the ability to acquire a fully malignant phenotype, to invade and infiltrate distant tissues. Our model predicts that each cancer harbors mutations in one or more MP pathways and/or one or more GF pathways, and the number of damaged pathways be proportional to the phenotypic ``distance'' of the tumour from its normal counterpart, i.e. inversely proportional to the tumour grade.  
 
\colorbox{colmb}{Based on}studies of skin cancer, the process of carcinogenesis is traditionally divided into three phases: initiation (induction of permanent alterations to DNA), promotion (proliferation of the initiated cell triggered by subsequent stimuli) and progression (from a benign tumour to a malignant one). Our interpretation is that damages to MP and GF pathways correspond to initiation, the first stage of carcinogenesis. The analogous in our model of the promotion phase, i.e. the proliferation of the stem cell, is postponed until regeneration occurs, or until the clock reaches the timer value. The action of a timer-dependent mechanism, if present in biology, may explain the long latency period observed between the exposure to carcinogenic substances (e.g. tobacco smoke) and the appearance of a tumour (e.g. lung cancer). Finally, progression, the third and last phase of carcinogenesis, may be linked to further mutations which confer additional powers to the already transformed cells, i.e. the capacity to infiltrate tissues and to produce distant metastases.

\colorbox{colmb}{Our proposal}is fully consistent with the cancer stem cell model, with the recent paradigm of dynamic stemness, and with the idea that each cell in a tumour has tumorigenic potential. This potential is associated to a damage to MP or GF genes, which is present initially in a single stem cell, and subsequently spreads to all normal and stem cells derived from it. The ET contribution to the field mainly lies in two ideas. The first idea consists in the notion of stem cell generation, which explains the emergence of a hierarchy of cell types, and brings together embryonic, adult and cancer stem cells under a unified framework. The second idea is a proposal for a low-level generation mechanism, through the actions of morphogens and growth factors.



\section{Conclusions}
  
We wish to conclude this summary indicating some experiments that could be used to test ET. One of the pillars of the model is the dynamic nature of the normal-stem transition. Stem cells can generate normal cells, and normal cells can revert to stem cells, with a new mobile code. This provides a conceptual bridge that links together embryonic, adult, ageing and cancer stem cells, and explains the emergence of a hierarchy of stem cells, a fact corroborated by data on numerous organs or biological subsystems. In order to validate the model, it is necessary to reproduce the biochemical mechanism which enables the normal-stem conversion, and shed light on the interplay between MP pathways and Hox genes, which is deemed responsible for mobile code generation.
  
As we have seen in section \ref{sec:evo-devo}, germline penetration implements a flow of genetic information from somatic to germline cells, to be passed on to future generations: as such, it can be considered the carrier of a transposon-mediated inheritance device. As a first step to prove the existence of the flow, we need to identify the moment in which transposable elements exit their host stem cells to reach the circulatory system, and make their way towards the germline. Furthermore, our model predicts that the spread of transposons is an event which immediately (in evolutionary terms) follows a lineage change. This could be done through techniques for estimating the age of DNA sequences more sentitive than those currently in use. 

In section \ref{sec:cancer} we have proposed an explanation for the heterogeneity of patterns of gene mutations observed in different tumours. The explanation is based on a hypothetical differential involvement of MP and GF pathways in the generation (and regeneration) of different organs. This prediction is susceptible of verification with bioinformatical techniques. To achieve this objective, MP and GF genes should be screened for their mutation rates associated to various types of cancer.

In conclusion, we have shown that ET is able to generate arbitrary 3-dimensional cellular structures of unprecedented complexity, and can reproduce a simplified version of some key biological phenomena such as development, the presence of transposable elements in the genome, ageing and cancer. For this reason, we think it can be proposed as a general-purpose model for multicellular biology. Future work will be aimed at further reducing the biological gap, mapping the model variables to individual genes and chemical elements.

\singlespacing
\bibliographystyle{plain}
\bibliography{ldanxmodst}
 
\end{document}